\documentclass{appolb}
\usepackage{graphicx}
\usepackage{subcaption}
\usepackage{cite}
\usepackage[breaklinks,colorlinks,urlcolor=blue,citecolor=blue,linkcolor=blue]{hyperref}

\begin{document}
% \eqsec  % uncomment this line to get equations numbered by (sec.num)
\title{QCD in the cores of neutron stars%
\thanks{Presented at Quark Matter 2022}%
% you can use '\\' to break lines
}
\author{Oleg Komoltsev,
\address{Faculty of Science and Technology, University of Stavanger, 4036 Stavanger, Norway}
%\\[3mm]
%{Third Author % of different affiliation
%\address{affiliation}
%}
%\\[3mm]
%the Name(s) of other Author(s)
%\address{affiliation}
}
\maketitle
\begin{abstract}
I discuss why state-of-the art perturbative QCD calculations of the equation of state at large chemical potential that are reliable at asymptotically high densities constrain the same equation of state at neutron-star densities. I describe how these theoretical calculations affect the EOS at lower density. I argue that the ab-initio calculations in QCD offer significant information about the equation of state of the neutron-star matter, which is complementary to the current astrophysical observations.
\end{abstract}
  
\section{Introduction}
The equation of state (EOS) of the dense matter at zero temperature is a necessary input for the neutron-stars (NS) physics. Theoretical calculations of the EOS can be done only at the two opposite (low- and high-density) limits. At the low-density limit the matter can be described within the chiral effective field theory (CET) \cite{Tews:2012fj,Drischler:2017wtt}. Those calculations are reliable up to around nuclear saturation density  $n_s = 0.16/\textrm{fm}^3$. On the other hand we can access the EOS using perturbative Quantum Chromodynamics (pQCD) at the asymptotically high densities, above $\sim 40n_s$ \cite{Gorda:2018gpy,Gorda:2021znl}. Central densities of maximally massive neutron stars are around $4-8 n_s$, which is not reachable within CET or pQCD. Therefore, there are no tools in our possession to compute EOS of the cores of NS from the first principles.

However, we can obtain an empirical access to the cores of NSs using recent astrophysical observations. The most important probes of NS physics are the discovery of massive NSs \cite{Demorest:2010bx, Antoniadis:2013pzd,Fonseca:2021wxt}, mass - radius measurements \cite{Miller:2021qha,Riley:2021pdl}, and the gravitational-wave and multi-messenger astronomy \cite{TheLIGOScientific:2017qsa,GBM:2017lvd}. Utilizing all constraints coming from astrophysical observation as well as first principle calculations narrows down dramatically the range of possible EOSs, which allows us to use the densest objects in the Universe to test independently various beyond standard model scenarios and/or general relativity.

Majority of the EOS studies extrapolate CET EOS up to NS densities 5-10$n_s$ and conditioning it with the observational inputs. The results differ from the works that include high-density limit and interpolate between two orders of magnitude. The qualitative difference is in the softening of the EOS happening around $\epsilon \sim$750 MeV/fm$^{-3}$, which can be interpreted as quark matter cores inside the most massive NS \cite{Annala:2019puf}. 

In this work I answer why and how pQCD input offers significant information about the EOS at NS densities. I find that pQCD input propagates non-trivial constraints all the way down to 2.2$n_s$ just by using solely thermodynamic stability, consistency and causality \cite{Komoltsev:2021jzg}. In addition the complementariness of the pQCD input to the astrophysical observations was studied in \cite{Gorda:2022jvk}. I show that pQCD is responsible for the softening of the EOS at the NS densities. Therefore, it is essential to include pQCD input in any inference study of the EOS.

\section{Setup}

\begin{figure}[h!]
\centering
\begin{subfigure}{.49\textwidth}
  \centering
  \includegraphics[width=.87\linewidth]{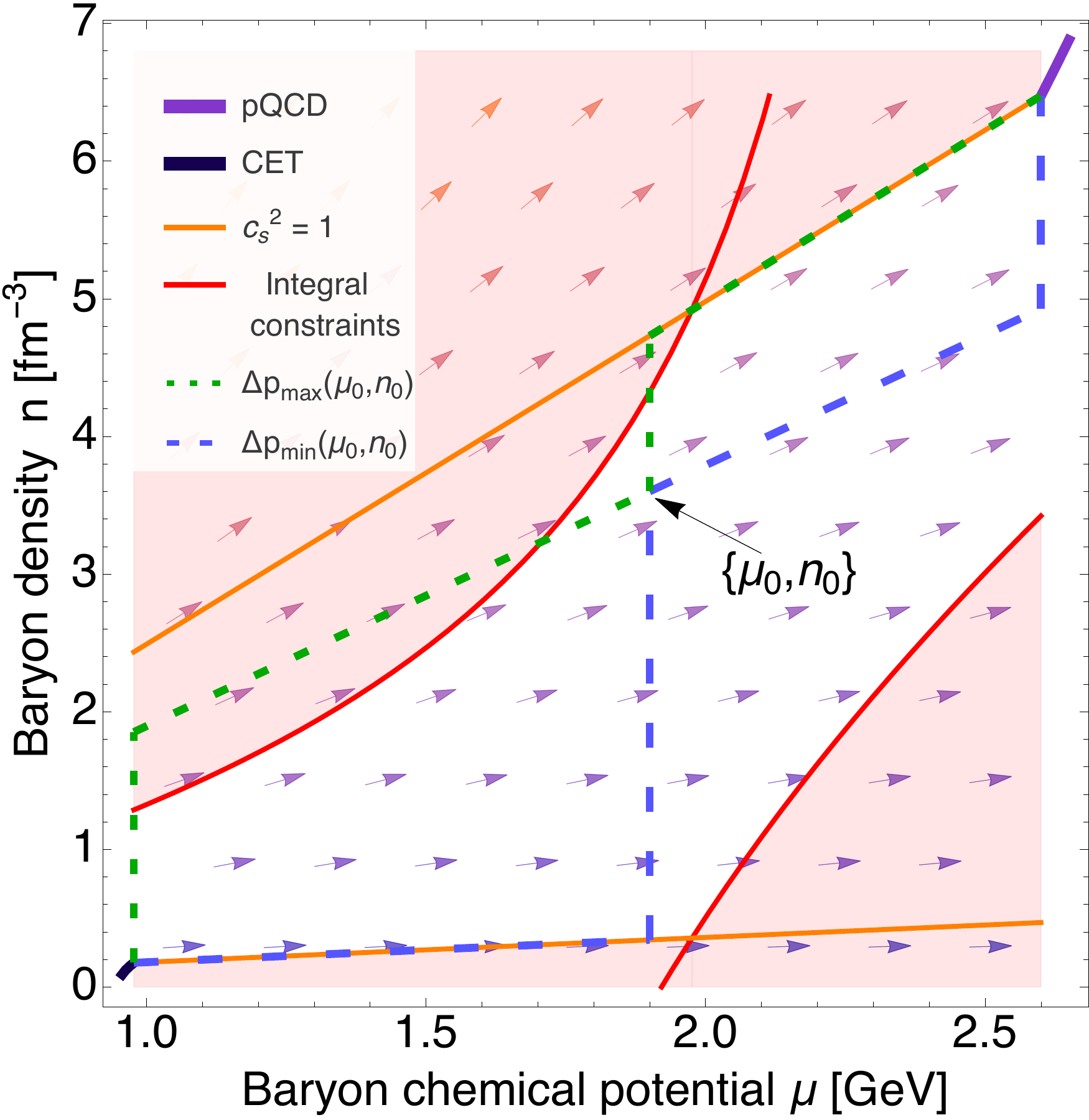}
  \caption{}
  \label{Fig:n_mu}
\end{subfigure}
\begin{subfigure}{.49\textwidth}
  \centering
  \includegraphics[width=.99\linewidth]{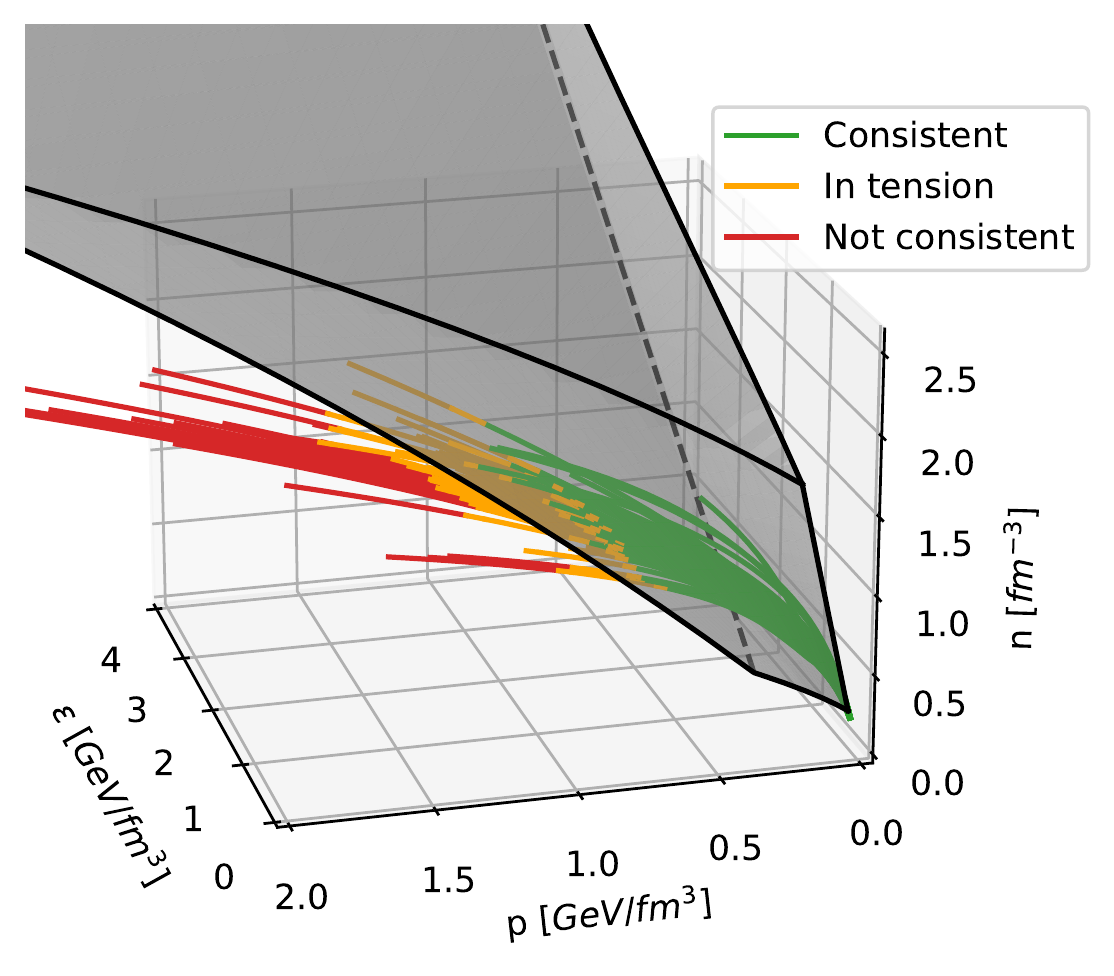}
  \caption{}
  \label{fig:1b}
\end{subfigure}
\caption{(\textbf{a}) - Baryon density as a function of chemical potential. Simultaneous fulfilment of thermodynamic consistency, stability and causality narrows down the allowed region (white area) of the EOS. (\textbf{b}) -   Zero temperature EOSs from CompOSE database plotted with the allowed area (gray shape) arising from the new constraints in $\epsilon - p - n$ space. Consistent/in tension/not consistent means that EOS is consistent with integral constraints for all/some/none X values in a range [1,4].}
\label{fig:1}
\end{figure}

All technical details as well as analytical formulas are presented in \cite{Komoltsev:2021jzg}. In this section I describe the conditions I use, in particular stability, consistency and causality, and the resulting propagation of the pQCD input down to lower densities. Let us start with the baryon density $n$ as a function of the chemical potential $\mu$ as shown in fig.\ref{Fig:n_mu}. The goal is to find all possible lines that connect endpoint of CET results (dark blue line in the bottom left corner) with the first point of pQCD calculations (purple line in the upper right corner) using 3 conditions.

%\begin{figure}[htb]
%\centerline{%
%\includegraphics[width=7 cm]{nvsmu_ver2.pdf}}
%\caption{Baryon density as a function of chemical potential. Simultaneous fulfilment of thermodynamic consistency, stability and causality narrows down the allowed region (white area) of the EOS.}
%\label{Fig:n_mu}
%\end{figure}

The first condition is thermodynamic stability, which implies concavity of the grand canonical potential $\partial^2_{\mu} \Omega(\mu) \leq 0$. At zero temperature $\Omega (\mu) = - p(\mu)$, which implies that the number density is monotonically increasing function of the chemical potential $\partial_{\mu} n(\mu) \geq 0$.

The second condition is causality -- the sound speed cannot exceed the speed of light $c^2_s \leq 1$. This provides constraints on the first derivative of the number density with respect to the chemical potential
\begin{equation}
    c^{-2}_s = \frac{\mu}{n}\frac{\partial n}{\partial \mu} \leq 1.
\end{equation}
For each point on the $\mu - n$ plane we can calculate the least allowed slope coming from causality, which is represented by the arrows in the fig.\ref{Fig:n_mu}. This cuts upper (lower) region of the plane, because any points from the area above (below) orange line $c^2_s=1$ cannot be connected to pQCD (CET) in a casual way. 

The third condition is thermodynamic consistency. In addition to $n$ and $\mu$ we need to match pressure $p$ at the low- and high- density limits. The pressure is giving by the integral of the number density
\begin{equation}
    \int^{\mu_{\rm QCD}}_{\mu_{\rm CET}} n(\mu) d\mu = p_{\rm QCD} - p_{\rm CET} = \Delta p.
\end{equation}
This implies that the area under the curve for any EOS is fixed by our input parameters. For each arbitrary point ${\mu_0,n_0}$ we can construct the EOS that maximize/minimize the area under the curve $\Delta p_{max/min}(\mu_0,n_0)$ shown as a green/blue dashed line in the fig.\ref{Fig:n_mu}. If $\Delta p_{max}(\mu_0,n_0) < \Delta p$ then any EOS that goes through the point ${\mu_0, n_0}$ does not have enough area under the curve. This discards the region in the lower right corner in the fig.\ref{Fig:n_mu} under the red line called "integral constraints". If $\Delta p_{min}(\mu_0,n_0) > \Delta p$ then any EOS that goes through the point ${\mu_0, n_0}$ has too much area under the curve. This cuts area in the upper left corner above the red line. The integral constraints can be obtained without any assumptions of interpolation function in a completely general and analytical way.

\begin{figure}[htb]
\centerline{%
\includegraphics[width=8.5 cm]{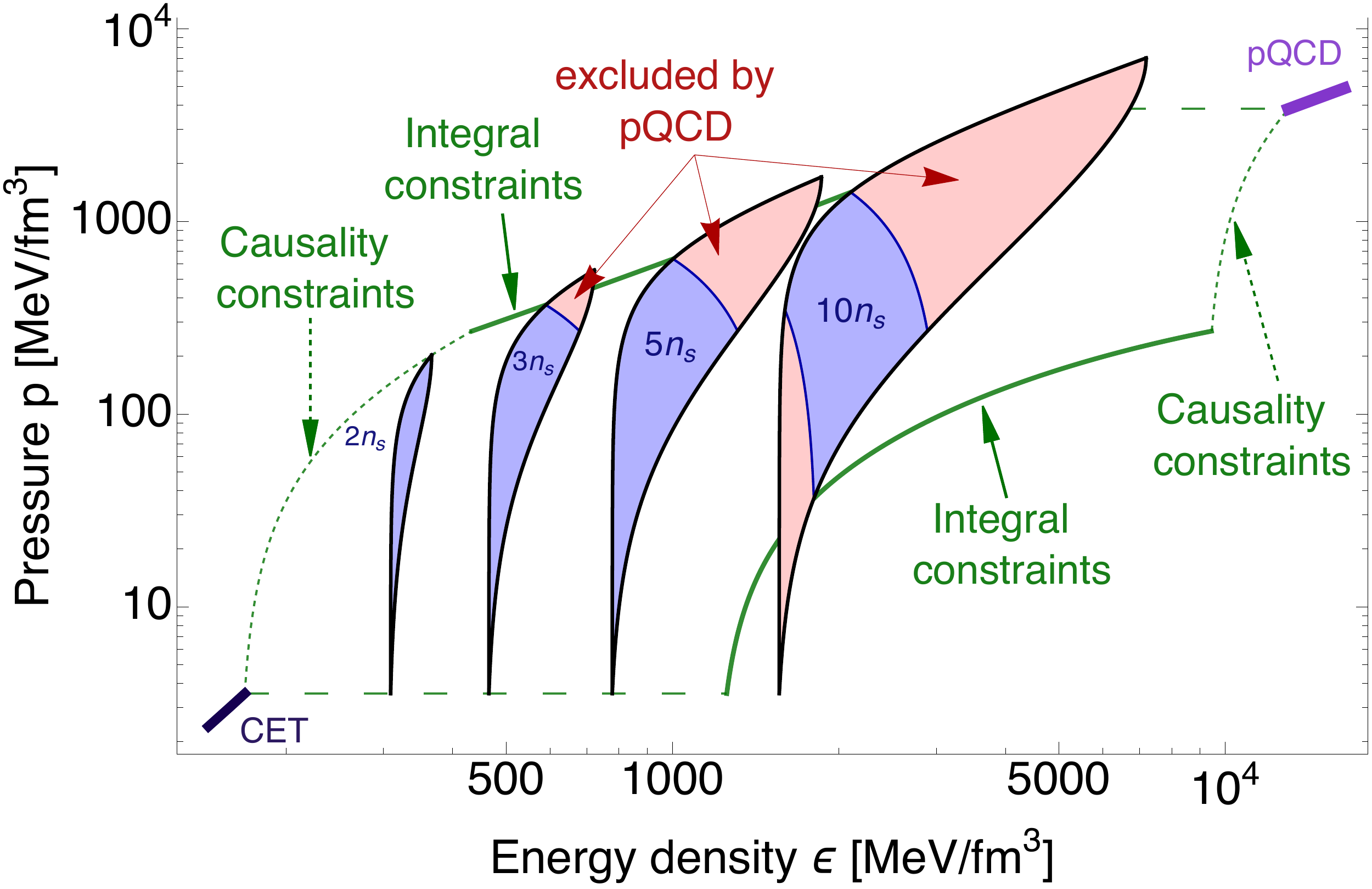}}
\caption{Constraints on the $\epsilon - p$ plane coming from low- and high-density limits. Shapes outlined by solid black line are the allowed areas for fixed number density without pQCD. Blue shapes are allowed regions after imposing pQCD input.}
\label{Fig:e_p}
\end{figure}

We can map the allowed region from $\mu-n$ to $\epsilon-p$ plane. The results of such mapping is shown in the fig.\ref{Fig:e_p}. The green envelope corresponds to the the white area in the fig.\ref{Fig:n_mu} restricted by the causality and the integral constraints. The shapes of allowed region with and without pQCD are shown for the fixed number density $n$ = 2,3,5 and 10$n_s$. This explicitly shows how pQCD input can propagate information down to lower density starting from 2.2$n_s$. And, strikingly, at 5$n_s$ it excludes 75\% of otherwise allowed area. 

Using the new constraints we can check the consistency of publicly available EOSs. Results for all zero temperature EoSs in $\beta$-equilibrium from the public CompOSE database \cite{Typel:2013rza, Oertel:2016bki} are shown in the fig.\ref{fig:1b}. Almost all of the EOSs start to be inconsistent with pQCD input at some density within the provided range. 

%\begin{figure}[htb]
%\centerline{%
%\includegraphics[width=7cm]{3D_database.pdf}}
%\caption{Zero temperature EOSs from ComOSE database plotted with the allowed area (gray shape) arising from the new constraints in $\epsilon - p - n$ space. Consistent/in tension/not consistent means that EOS is consistent with integral constraints for all/some/none X values in a range [1,4]. Almost all public EOSs are inconsistent with QCD input at some point in density.}
%\label{Fig:compose}
%\end{figure}

\section{Bayesian inference of EOS}

With the construction described above we can propagate information from ab-initio QCD calculations down to NS densities, where we already have constraints from astrophysical observations. To understand if the new constraints from pQCD go beyond the constraints coming from the NS measurements we construct a Bayesian-inference framework. This was done in \cite{Gorda:2021znl}, where we generate a large ensemble of different EOSs using Gaussian-process regression. We anchor the ensemble to CET calculations and extrapolate it up to 10$n_s$, where we impose pQCD input as a blue shape from fig.\ref{Fig:e_p}. We condition ensemble sequentially with the astrophysical observations. With this setup we can turn on and turn off pQCD input in order to study its effect on our posterior after imposing astrophysical observation.

\begin{figure}[htb]
\centerline{%
\includegraphics[width=10.5cm]{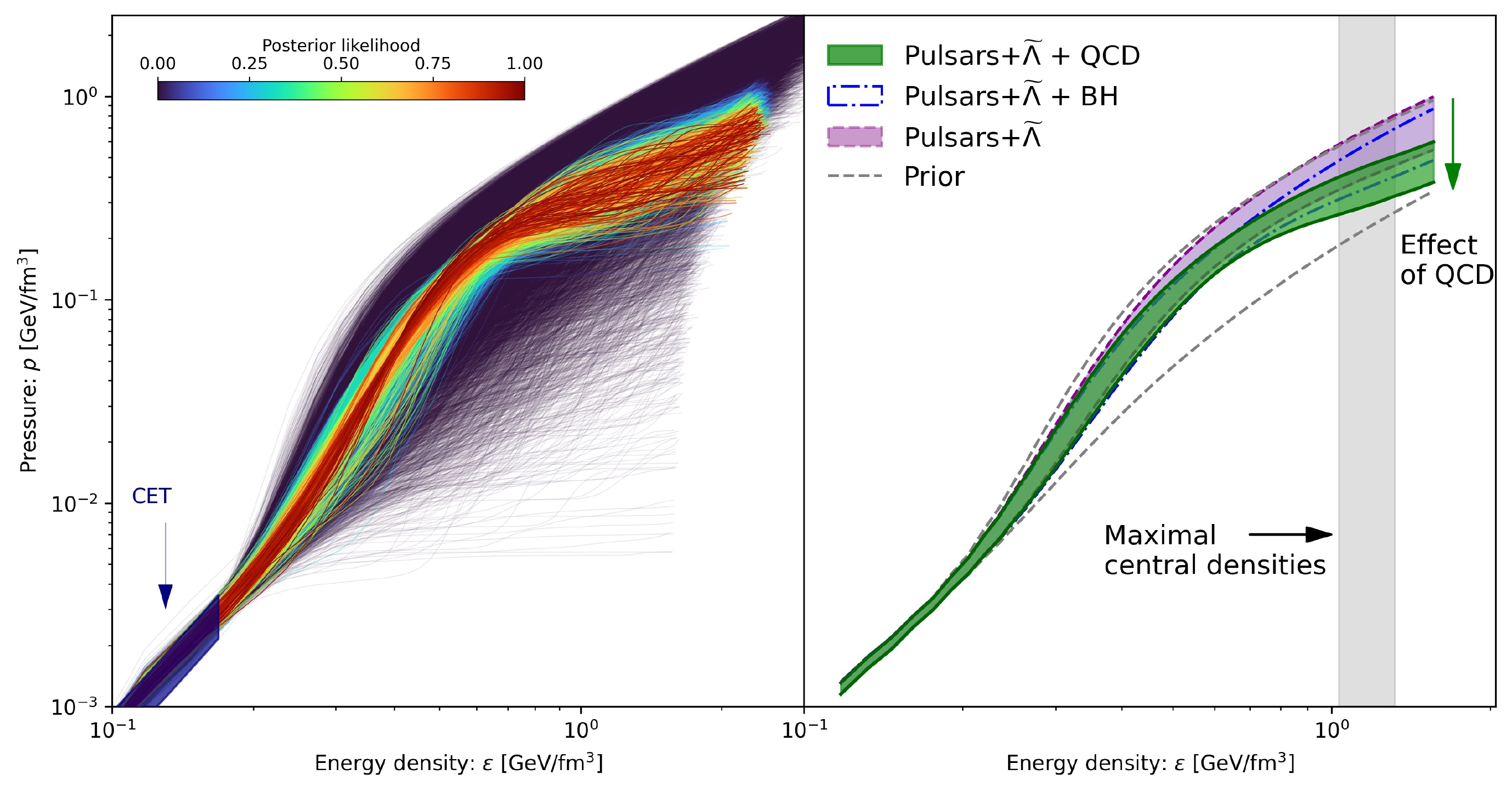}}
\caption{Left plot shows the sample of 10k EOSs. The coloring represents the likelihood after imposing all observations as well as pQCD input. Right plot shows 67 \%-credible intervals conditioned with the different astrophysical observations and high-density limit. The gray band shows 67 \%-credible interval for the maximal central energies density reached in NSs.}
\label{Fig:gp}
\end{figure}

The results are present in fig.\ref{Fig:gp}. The reduction of the pressure (green arrow on the right plot), which is caused by the QCD input, happens before the density reaches its maximal central value. In another words, the prediction of QCD input is the softening of the EOS that happens inside the most massive neutron stars.
\section{Conclusion}

In this work, I show how QCD calculations at asymptotically high densities can propagate information down to lower densities using solely thermodynamic consistency, stability and causality. This information offers a significant constraints to the EOS at NS density, which is complementary to the current astrophysical observations. In addition, I show that the prediction of QCD input is the softening of the EOS that happens in the most massive NSs. An easy-to-use python script is provided to check consistency of the EOS with pQCD input, available on \href{https://github.com/OKomoltsev/QCD-likelihood-function}{Github} \cite{OlegGithub}. 

In order to achieve accurate determination of the EOS it is crucial to utilize all available controlled measurements and theoretical calculations. This strategy either helps us to understand the matter of the densest objects in the Universe or find a discrepancy between different inputs, which allows us to use NS as a tool for fundamental discoveries. 

\bibliographystyle{IEEEtran}
\bibliography{main.bib}

\end{document}